\documentclass[a4paper,11pt]{article}
\pdfoutput=1 

\usepackage{jcappub} 

\usepackage[T1]{fontenc} 

\title{Generation of UL-Axions by a Kerr Black Hole. Part III: The fate of its Axion Wind}

\author{G. Pelletier}
\affiliation{Univ. Grenoble Alpes, CNRS, IPAG, 38000 Grenoble, France}

\emailAdd{astroguypel@gmail.com}

\abstract{In two previous articles, we have explained how super-massive Kerr Black Holes in the nucleus of an active galaxy or quasar can generate a large amount of axion particles in their magnetosphere. In that third part, we explore the possibility of observable dynamics resulting from some interaction of an axion wind with the baryonic matter and the axionic Dark Matter in the surroundings of the closest AGNs. The axions that we have considered are the same as those Ultra Light Axions favored in cosmology as the main content of Dark Matter and which form Bose-Einstein Condensates, supporting a gigantic Schr\"odinger wave. In the paper we investigate what can be the nature of the interaction of the axion wind with the plasma and the hydrogen gas in the host galaxy and its halo. Despite the intriguing nature of a possible interaction, we show that only a kind of gravitational Tcherenkov-Vavilov effect can be efficiently operating with acoustic waves through a streaming instability. The instability is  fast enough to operate efficiently in the central region of the host galaxy up to parsec scales and makes Lyman-alpha diagnostics of the acoustic waves of hydrogen gas and plasma possible, as well as the detection its hyperfine emission at 21 cm. However it turns out that scattering of the axion wind is very long and thus these winds could travel over very large distances and cross other galaxies. And then the wind travels without significant loss along its halo and even beyond up to quantum thermalization, which makes possible direct detection of these axions by conversion into gamma-rays on Earth, opening a possible axion astronomy in the local Universe. The parameters of Messier 87 indicate that it is a good candidate for developing these investigations both through the conversion of axions into gamma-rays and Lyman alpha diagnosis in the bulge. }

\begin{document}
\maketitle
\flushbottom

\section{Introduction}

Since three decades the notion of axion particle has been designed in order to solve the problem of anomalies in QCD theory (Peccei \& Quin 1977 \cite{PQ}). However outside particle physics the impact has remained poor. Two important publications renewed the interest of the concept of Axion particle, one \cite{HOTW} in 2017, by renown scientists who argued the
advantages of axions, even Ultra Light Axions (ULA), in cosmology; another one in 2020, by P. Sikivie who published an important review \cite{SP1} covering the basic
physical developments of the concept of axions, as well as the experimental methods envisaged for their detection. Besides the notion of axions having a mass $\sim 10^{-6} eV$, designed to solve the anomalies of the QCD theory causing the CP violation, a renewal raised with the publication of Svrcek \& Witten (\cite{SW} 2006) who have shown the interest of this mechanism to solve the problem of anomalies in string theory with a wide range of axion masses and notably ultra light masses. So  an interest for the ULA, Ultra Light Axions, has been growing, especially in cosmology, as a promising candidate for Dark Matter. In a recent publication, D. Marsh \cite{DM} has developed a convincing cosmological model of Dark Matter involving ULA (also called Fuzzy Light  Particles)
of masses in the range ($10^{-20} eV - 10^{-18} eV$), the upper bound being a constrain to avoid the development of a closed Universe, the lower bound being a constrain imposed by Lyman-alpha observations. The primordial generation of these cosmological ULA is produced by the initial re-alignment of the axion field in the potential well responsible for the symmetry breaking that allows to cancel the anomaly.

In this paper we continue the analysis of the production of these ULAs in the magnetosphere of a Kerr Black Hole (KBH), that we recently published  (\cite{GP1} and \cite{GP2}).
In these papers we have shown that a super-massive Kerr Black Hole (KBH ) is capable to generate a large amount of ULAs in Active Galactic Nuclei (AGNs), forming a relativistic wind, whose luminosity is a small but sensitive fraction of the Blandford-Znajek \cite{BZ} luminosity. The sections 2 and 3 of this paper summarize these investigations developed in Papers 1 and 2, that concerns the source of ULAs in AGNs and brought supplementary details important for the development of this third part that mostly concerns the physics of the ULA relativistic wind itself and its observability. Most efforts for the detection of ULAs in astrophysics has been focused on the conversion process of axions into $\gamma$-photons with a promising attempt with a close Active Galactic Nucleus, namely M87. The investigators \cite{MEX} obtained a suggesting result with a significance of $\sim 1.56 \sigma$; the authors hopefully expect a significance reaching $\sim 3 \sigma$, if future gamma ray observations allow to push the flux upper limits at tens of TeV.

Those axions particles are {\it pseudo} scalars, because they are created through $\vec E \cdot \vec B \neq 0$ which violates parity. Whereas $\vec E \cdot \vec B = 0$ almost everywhere in astrophysics, there are some magnetospheric sheets where $\vec E \cdot \vec B \neq 0$ and axions are created and ejected.
Indeed two mechanisms of ejection are at work : the formation of both backwards and forwards axion waves in an oscillating ``gap" and also a participation of the axions in the mechanism of Blandford-Znajek \cite{BZ}. A gap is a thin layer of electrons and positrons around the Black Hole horizon or a neutron star that contains a longitudinal component of electric field (i.e. $\vec E \cdot \vec B \neq 0$) which is often unstable and oscillates due to the unavoidable constitutive conflict between the creation of $e^+e^-$-pairs and their screening of the longitudinal electric field which contributes to create them (see Timokhin \& Arons 2012 \cite{TA} in the case of Pulsars, Crinquand et al. 2020 \cite{BC}, Pelletier 2025 \cite{GP2} in the case of Kerr Black Holes). That  inherent conflict is summarized in section 4.

The creation of ULAs in oscillating gaps is so efficient that they form Bose-Einstein Condensates (BEC) even with our ``preferred" ULA mass of  $10^{-18} eV$, because their Compton length $\ell_c$ is comparable with the gravitational radius of massive KBH ($r_g \sim 10^{13} cm \sim 10^{-5} pc$). That range is most favorable for having efficient ULA luminosity, for a magnetic field of $10^4 G$ on the Black Hole horizon. Nevertheless the process involved in the theory also works for lower axion masses down to $10^{20} eV$, especially for a higher Black Hole mass. A discussion with varying axion mass, Black Hole mass and magnetic field intensity is proposed, including the possibility of a so called ``MAD" regime of the accretion disk where the magnetic field could be $50$ times larger than its equipartition value on the horizon \cite{TCH} \cite{NAR}.

As will be shown in this paper, a significant relativistic wind of ULA is produced in this mass range of cosmological interest. This means that a secondary production of Dark Matter can be expected from the AGNs, but not copious enough to increase significantly $\Omega_{DM}$ which remains in the error bar of Planck measurements, because the amount of AGNs is only $20 \%$ of the galaxy population.
The abundance of created axions is estimate in section 5 and the formation of Bose-Einstein Condensates (BEC) is emphasized, which leads to describe the axion wind by a single Schr\"odinger wave function in its co-moving frame.

In section 6, we present again the hydrodynamical aspect of that gigantic Schr\"o- dinger wave under the influence of gravity \cite{HOTW} and discuss the issue of the relaxation of that unusual hydrodynamics through quantum interferences. It turns out that quantum thermalization occurs at very large distances, several $100 \, Mpc$.

These so light bosons having no charge of any kind can only undergo collisionless gravitational interaction that turns out to be weak. The strongest interaction occurs through a kind of gravitational Tcherenkov-Vavilov effect that generates acoustic waves which leads to a strong instability in $G^{1/2}$, due to the back reaction of acoustic waves on the wind through the gravitation field. This is presented in section 7. That generation of acoustic waves occurs at parsec scales in the centre of the host galaxy and offers the possibility of Lyman-$\alpha$ and hyperfine hydrogen lines et 21 cm investigations. As for the expected scattering of the axions off these waves, it remains very weak and then the axion wind continue its travel across the halo and the surrounding extragalactic medium, which opens possible  detections of extragalactic axions on Earth and an axion astronomy. A summary of the results of this study is proposed in a conclusive discussion in section 8.

\section{Ultra Light Axions and massive Kerr Black Holes}

Axions are characterized by two parameters : their mass $m_a$ and their ``decay constant" or symmetry breaking energy : $f_a$.
Their product is related to the amplitude of the potential well $\Lambda$ such that $\Lambda^2 \equiv \xi m_a f_a$.

The strength of the interaction of axions with an electromagnetic field is measured by a
coupling parameter
$$
 g_{a\gamma} = \xi \alpha_{e.m.}/f_a \, .
 $$
The dynamics stems from the Lagrangian :
$$
{\cal L} = {\cal L}_a + g_{a\gamma} a \vec E\cdot \vec B -{1\over 4}(\vec E^2- \vec B^2) \ .
$$
Whereas the Lagrangian for the electromagnetic interaction involves the first Lorentz invariant only, the interaction with axions involves the second Lorentz invariant, which breaks the parity
symmetry. The axions potential  $V(a) = {1\over 2} {a^2 / \ell_c^2}$ (at leading order), where the axion Compton length $\ell_c \equiv \hbar/m_ac$ that turns out to be very large for very small mass.\\

{\it The Kerr Black Hole parameters} : \\

Its mass $M_*$ determines the gravitation radius $r_g \equiv GM_*/c^2$; the ``spin" parameter $a_s$ (where $0 \leq a_s \leq 1$ is ) determines the rotation velocity $\Omega_h= a_s r_h/c$  on the Black Hole horizon. There are three surfaces: the horizon $r_h = r_g(1+\sqrt{1-a_s^2})$, the ergosphere : $r_e = r_g(1+ \sqrt{1-a_s^2 \cos^2 \theta})$ and the stagnation surface, similar to a sphere a little ``battered", $r_s(\theta)$ which is a zero of a 4th degree polynomial varying between $1.6 r_g$ and $2r_g$).
Such a rotating Black Hole supports a rotating magnetosphere of angular velocity $\Omega_* < \Omega_h$. The intensity of the magnetic field on the horizon $B_*$ is commonly considered to be close to the equipartition estimate obtained in the inner region of the accretion disk $B_* = 10^{4} \left(\frac{10^8 M_{\odot}}{M_*}\right) \, Gauss$. However it has been recently shown \cite{TCH} \cite{NAR} that a Black Hole can support a magnetic field almost fifty times larger on its horizon.

A Poynting flux generated by a Kerr Black Hole endowed with a magnetosphere is measured in term of the fiducial power :
\begin{equation}
\label{L*}
L_* \equiv cB_*^2 r_g^2 \simeq 3 \times 10^{44} erg/sec \ .
\end{equation}
These values of the parameters characterized the most common KBH in AGNs; we will call this references as the fiducial ones and will further discuss some deviations from this set of references.\\

{\it Parameters of the considered Ultra Light Axions} :\\

There is a reference value $a_*$ for the energy achieved by the axion waves created by a longitudinal electric field by defining
$a_*$ as an energy achieved by an axion wave by a longitudinal static electric field of intensity $E_* = -B_*$; and one obtains
\begin{equation}
\label{A*}
a_* = g_{a\gamma} \hbar c \ell_c^2 B_*^2 \ .
\end{equation}

A reference axion luminosity  can be derived from that expression of $a_*$, this is
\begin{equation}
\label{P*}
P_* \equiv a_*^2/\hbar \ .
\end{equation}
And it turns out that this power can be expressed in term of $L_*$ :
\begin{equation}
\label{PL*}
P_* = g_{eff} {\ell_c^2 \over r_g^2} L_* \ ,
\end{equation}
where the non-dimensional coupling constant
\begin{equation}
\label{GF1}
g_{eff} \equiv g_{a\gamma} a_* = \alpha_{e.m.}^2 \frac{(\hbar c)^3}{\Lambda^4} B_*^2 \ .
\end{equation}
Numerical estimate :
\begin{equation}
\label{GF2}
g_{eff} \sim \frac{22 \, (eV)^4}{\Lambda^4} \left(\frac{B_* }{10^4 \, G} \right)^2 \ .
\end{equation}
Therefore $g_{eff} \sim 1$ can be expected for $\Lambda^2 \sim 5 \, (eV)^2$, so that for $m_a \sim 10^{-18} eV$, $f_a \sim 5 \times 10^{18} eV$, which is acceptable for cosmology purpose, but excludes QCD axions for which $\Lambda^2 \sim 10^5-10^6 eV^2$ even with the most intense magnetic field generated in Nature, i.e. $B = 10^{13} G$ . We note that for the lowest value of $m_a$, which is $10^{-20}eV$ according to $Lyman \, \alpha$-forest measurements, $g_{eff}$ is still of order unity with $f_a < 5 \times 10^{20}eV$; it is considered that the Universe is closed if the cosmological axions have a “decay constant” $f_a > 3 \times 10^{20}eV$.

Kerr Black Holes of $M_* \sim 10^8 M_{\odot}$ with a magnetosphere having a magnetic field $B_* \sim 10^4 \, Gauss$ received a strong consideration because they are considered as generating relativistic jets of luminosity $\sim 10^{44} \, erg/s$, a sizable fraction of $L_*$. Their gravitation radius $r_g \sim 10^{13} \, cm$. It turns out that the Compton length of axions of mass $m_a \sim 10^{-18} \, eV$ is comparable with the gravitational radius. Those masses correspond to the upper limit of cosmological axions, which have a mass constrained to be in the intervalle ($10^{-20}-10^{-18} \, eV$). Axions of mass significantly larger than $10^{-18} \, eV$ are not efficiently generated by Kerr Black Hole; moreover they are not considered as candidates for the Dark Matter constituents because they would close the universe. Axions of mass significantly smaller than $10^{-18} \, eV$ are considered as forming a Bose-Einstein condensate wrapping around the Black Hole; the axions of mass close to $10^{-18} eV$ are also forming Bose-Einstein Condensates (BEC). If the intensity of the magnetic field on the KBH horizon  can reach $50 \times B_*$, then all these axions are copiously generated by the KBH. Those cosmological axions are not those associated with QCD, which lead to a very small value of $g_{eff}$.

At this stage, it is worth to look at the effects of changing the values of the important parameters, $B_*$ and $m_a$, without changing the Black Hole mass $M_* = 10^8 M_{\odot}$, and $\Lambda^2 = \xi m_a f_a \sim a \, few \, (eV)^2$. Fixing $M_*\sim 10^8 M_{\odot}$ implies that one considers the maximum intensity of the equipartition B-field achieved in the inner region of the accretion disk. However it has been recently shown that the intensity of the field incorporated in the Black Hole can be $50$ times larger \cite{TCH}, \cite{NAR}. As for the axion mass, the value of  $10^{-18} \, eV$ has been privileged in both previous papers \cite{GP1} and \cite{GP2}, because it leads to  a Compton length comparable with the gravitational radius of a KBH of $10^8 M_{\odot}$ in an Active Galactic Nucleus; however the Ultra Light Axions considered in cosmology have a mass between $10^{-20}- 10^{-18} eV$.
The main scalings are the following : $g_{eff} \propto B_*^2$, we will see latter on that the axion luminosity $L_a = \eta_a L_*$, with $\eta_a = g_{eff} \chi_0^2 (h^2/\ell_c^2) \propto m_a^2 B_*^2$ and thus $L_a \propto m_a^2 B_*^4$. Three other important scalings will be presented later on.

However under this choice of reference power $P_*$ of axion emission by these sources is close to their jet luminosity $L_*$.
In fact it will turn to be a smaller fraction of $P_*$, because the generation of axions occurs in gaps of thickness $h$ significantly smaller than $r_g$, which leads to an axion  wind luminosity quite smaller than the jet luminosity.\\

\section{ULA Generation by an electric field such that ($\vec E \cdot \vec B \neq 0$)}

The generation of axions is due to a non-vanishing second Lorentz invariant of the electromagnetic field $\vec E \cdot \vec B$, especially when it undergoes oscillations. This is formalized by a source term in the Klein-Gordon equation:
$$
g_{\mu \nu} \partial^{\mu} \partial^{\nu} a + {\partial \over \partial a} V(a) = - g_{a\gamma} \hbar c \vec E \cdot \vec B \ ,
$$
where the potential for the axion field is $V(a) = {1/2} (a^2/\ell_a^2)$ at leading order.
When $\vec E \cdot \vec B$ is static, or has a static component, a static axion wave is generated. For a quasi-static ($\omega \ll \omega_a$) and long wavelength excitation ($\lambda \gg \ell_a$),
$$
\bar a \simeq -g_{a \gamma} \hbar c \ell_c^2 \vec E \cdot \vec B \ .
$$
This is the highest value $a_*$ that can be achieved by the axion field, especially for $\vec E = - \vec B_*$.

Let us now consider an oscillating electric field aligned with the a static magnetic field : $\vec E = - \chi(x,t) \vec B_*$, located in a thin sheet  of width $h$, in the form of a disk of radius $R$,
orthogonal to $\vec B_*$. For simplicity, assume $\chi(x,t) = \chi_0 \psi(x) \cos \omega_0t$, with $\psi(x) = \sin(\pi x/h)$. Then, in a flat space, the Klein-Gordon equation becomes an equation for a driven harmonic oscillator:
$$
\ddot a + \omega^2 a = \omega_a^2 a_* \chi_0 \cos \omega_0t \ ,
$$
where $\omega^2 = \omega_a^2 + c^2/h^2$. A resonance occurs for $\omega = \omega_0$. That resonant classical solution then is :
$$
a(x,t) = {1\over 2}a_* \chi_0 {\omega_a^2 \over \omega_0^2} \omega_0t \sin \omega_0t \sin(\pi {x \over h}) \ .
$$

However, creation of axion through that excitation would require a quantum description. Let us see if the quantum solution will significantly differ from the classical one.\\

{\it The quantum approach}\\

Here we just recall the useful results derived in \cite{GP1}.
Selecting a resonant mode, the quantum description involving annihilation and creation operators, $\hat A, \hat A^{\dag}$, is quite simple.
$[\hat A, \hat A^{\dag}] = 1$, $[\hat q, \hat p] = i\hbar$, where $a = q a_*$ and $[\hat q, \hat {\dot q}] = i\nu$. The momentum operator $p$ differs from $\dot q$ by a factor $\kappa$. The frequency $\nu = \hbar/\kappa$;
whereas $\nu = \omega_a$ for free axions, it turns out that $\nu$ is much smaller than $\omega_a$ in the context of the macroscopic excitation. Indeed
\begin{equation}
\label{NU}
{\nu \over \omega_a} = \frac{m_ac^2}{g_{eff} L_* h/c} =  \frac{{8\pi} \, m_ac^2}{g_{eff} {\cal E}_{mag}} \ ,
\end{equation}
where $L_* = B_*^2 R^2 c$ is the flux of magnetic energy across the source and ${\cal E}_{mag} \equiv B_*^2 R^2h/8\pi$ is the magnetic energy in the source.
The amount of axion energy in the source is $$
{\cal E}_a = m_a c^2 {\omega_a \over \nu} {h^2 \over \ell_c^2} {\chi_0^2 \over 8} \ .`
$$
The average number of axions in the source is
\begin{equation}
\label{Na}
N_a = {\omega_a \over \nu} {h^2\over \ell_c^2} {\chi_0^2 \over 8} \ .
\end{equation}
Because $\nu/ \omega_a$ is very small, $N_a$ is a very large number ! And the power emission of axions is
\begin{equation}
\label{Q_a}
Q_a = g_{eff} L_* {\chi_0^2 \over 4} {h^2 \over \ell_c^2} \ .
\end{equation}
The quantification of oscillation states  leads to an incrementation of $\sqrt{\nu \omega_0} \ll \omega_0$  which is very small and the occupation number increases to a high value. The separation between quantum
states are very tight and the occupation number very large. Thus the dynamics is very close to the classical prediction, which is quite significant.

\section{The KBH magnetosphere as a source of axions}

\subsection{Generation of magnetospheric sheets where $\vec E\cdot \vec B \neq 0$}

A magnetosphere of a KBH is governed by a force free equilibrium, with ``gaps" which are local anomalies of the equilibrium: whereas
$\vec E \cdot \vec B = 0$, almost every where, there are some thin sheets where $\vec E \cdot \vec B \neq 0$ due to the unavoidable Goldreich-Julian charge density \cite{GJ} (close to the horizon
$\rho_{GJ} \simeq \Omega_h B_*/4\pi c$) since ${\rm div} \, \vec E_t \neq 0$, a generation of a parallel component $\vec E_{\parallel}$,
having an intensity $\sim 0.1 B_*$ is unavoidable. A generalization of that Goldreich-Julian
charge density in General Relativity has been calculated by A. Levinson \& B. Cerutti \cite{LC}.

That extra parallel component of the electric field can be steady or unsteady.

These gaps has been revealed by numerical simulations close to the horizon of the Black Hole (\cite{BC}.
Their width $h$ is much smaller than the gravitational radius $r_g$. Most of them are vibrating at a pulsation of order $c/h$.
A theory has been proposed (\cite{GP2}). The overall mechanism is the following :
\begin{itemize}
\item Suppose one starts with a localized electric field along the poloidal magnetic field, over a width h, in a quasi empty space containing $e^+ e^-$ pairs.
\item That electric field accelerates electrons and positrons in opposite directions.
\item These electrons and positrons generate gamma rays by inverse Compton scattering off UV-photons emitted by the accretion disk.
\item Above some threshold, the encounters of gamma rays and UV-photons produce new $e^+ e^-$ pairs.
\item The abondance of pairs replenishes the gap and thus screens the electric field, which stops the pair catastrophe.
\item Pairs escape from the gap and the longitudinal electric field grows again because of the presence of the Goldreich-Julian \cite{GJ} charge density.
\end{itemize}

The typical time-scale of this process is the Compton time scale $\tau_c \equiv \lambda/c$ where $\lambda$ is the mean free path of Compton scattering, namely : $\lambda \equiv \sigma_T/n_{UV}$. The oscillating regime of gaps is obtained for a gap width $h = \lambda$ with a pulsation $\Omega_0 \sim c/\lambda$. It has been shown \cite{GP2} that the
width h is equal to the mean free path of the gamma photons scattering off the bath of UV photons, due to the creation of pairs $e^+e^-$.

Therefore one expects a generation of axions with an axion luminosity of the order of the previous calculation \cite{GP2}:
$$
L_a \sim g_{eff} \chi_0^2  {h^2 \over \ell_c^2} \, L_*  \ ,
$$
with $h \sim 10^{-2} r_g$ \cite{GH} for a single gap; however since there is a set of a dozen of gaps of that size between $r_g$ (but apart the horizon) and $r_0 \simeq 2r_g$ \cite{BC}, then we replace that value by an effective size of $0.1 r_g$. Now the frequency is not exactly the same in each gaps, because close to the horizon the measures of time and radial intervalle in the ZAMO frame is affected by the local value of the laps function $\alpha$ and the resonant frequency $\omega = \alpha \omega_0$; $\alpha(r)$ vanishes at the horizon and tends to 1 at large distance; at $2r_g$ $\alpha \simeq 1/2$.  Note that $\vec E \cdot \vec B$, that depends on $\sqrt{\vert g \vert}$, does not depend on $\alpha$. The size $h$ is changed in $h/\alpha$. Therefore a more careful calculation should incorporate these General Relativity scalings; for simplicity in this paper we took the opportunity to make them apart from the horizon in a local sheet of width much smaller than $r_g$ and consider the GR corrections as coefficients involving $0 < \alpha \lesssim1$ only.\\

Since we consider ULA having a mass such that $\ell_c \sim r_g$, the axion luminosity can reach $10^{-5}-10^{-4} L_*$. That emission seems weak, however it may have some interesting consequences in the AGN surroundings. An important remark : we started the analysis with the assumption that the Compton length should be close to the gravitational radius and thus $m_a \lesssim 10^{-18} eV$ seem to be the most appropriate axion mass. However a more accurate requirement is that $\ell_c \gtrsim \lambda = h$, thus for $\lambda$ significantly smaller than $r_g$, we can also select axions of mass $10^{-20}-10^{-19} eV$, which enlarges to the intervalle of masses relevant for cosmology and also increases the efficiency of the generation of axions, especially when $B_*$ is a few tens larger than the equipartition value.\\

However a fraction of the generated axions could be converted in X-rays and $\gamma$-rays in the source (see later on in section 6).

\subsection{Axions and the Blandford-Znajek process}

A Kerr Black Hole is characterized by having a surrounding space-time with a rotating frame in which the metric is independent of time; relatively to that specific frame (the so-called ZAMO frame) where the metric is stationary, the angular velocity is $\omega(r, \theta)$, independent of $\theta$ on the horizon where the laps function $\alpha(r, \theta) \equiv ds/dt$ vanishes ($\alpha(r_h) = 0$); the horizon is a sphere. The value $\Omega_h$ of $\omega$ on the horizon defines the angular velocity of the Black Hole.
At remote distance from the horizon, $\omega(r, \theta)$ declines like $r^{-3}$. There exists a surface, called the {\it ergosphere}, that envelopes  the horizon, where every particle is enforced to rotate at the same angular velocity than the Black Hole.
Kerr Black Holes are naturally imbedded by an axisymmetric magnetosphere that rotates also with an angular velocity $\Omega_* < \Omega_h$. Because the poloidal magnetic field threads the disk, an electromotive force develops, driving an electric current flowing about the axis towards the horizon :
$$
\vec E_t = \frac{\omega- \Omega_*}{\alpha c} \bar r \vec e_{\phi} \times \vec B_p \ .
$$
 A toroidal component of the magnetic field is thus generated:
 $$
 B_{\phi} = \frac{\omega-\Omega_*}{\alpha \beta_p c} \bar r B_p \ .
 $$
 ($\bar r$ is the radial distance to axis that differs from $r \sin \theta$ by GR-corrections).
 Close to the horizon, $\omega > \Omega_*$ and $\beta_p < 0$ (accretion), so that $B_{\phi} < 0$. So a braking torque is exerted on the horizon that behaves like a conductor. And thus a Poynting influx is produced close to the horizon; however it reverses beyond the stagnation surface, then becoming an out-flux. A careful investigation indicates that a Poynting out-flux is generated beyond a third surface, called the {\it stagnation surface} where the matter flow reverses from accretion to ejection; the regularity of the crossing of that surface requires that $\omega = \Omega_*$ on it. However the main energy out-flux results from a generalized magnetic Penrose effect \cite{RP} \cite{GPar} \cite{GPM} \cite{LGA}. Indeed because of the force free description (a highly satisfied approximation) the flux of angular momentum is dominated by the magnetic field. This is where axions modify the General Relativity electrodynamics, as opposed to a flat space-time; indeed one obtains an out-flux of angular momentum (\cite{GP2}) :
 $$
 \vec F_{\ell} = -{\bar r B_{\phi} \over 4\pi} (\vec B_p -g_{a\gamma} a\vec E_p) =  - {\bar r B_{\phi} \over 4\pi} (1+g_{a\gamma} <a\chi>) \vec B_p \ ,
 $$
 where we define the function $\chi(\vec x, t)$ such that $\vec E_p = - \chi \vec B_p$.
 The total energy flux is $\vec S_{tot} = \alpha^2 \vec S_{P} + \alpha \omega \vec F_{\ell}$, where $\vec S_P$ is the Poynting flux measured locally in the vicinity of the horizon.
 $$
 \vec S_P = {c\over 4\pi} \vec E_t \times \vec B_{\phi} = {1\over \alpha^2 } (\omega-\Omega_*)^2 \bar r^2 {B_p \vec B_p \over 4\pi \beta_p c} \ .
 $$
 That Poynting flux converges towards the horizon before the stagnation surface where $\beta_p <0$ and radiates outwards beyond that surface ($\beta_p > 0$).

 As for the contribution of the magnetic Penrose effect,
 $$
 \vec S_{\omega} = \omega \vec F_{\ell} = - {1\over \alpha}(1+g_{a\gamma}<a\chi>) \, \omega \, (\omega -\Omega_*)\bar r^2 {B_p \vec B_p  \over 4\pi \beta_p c} \ .
 $$
 This is an out-flux everywhere.
 Therefore the total energy flux
 $$
 \vec S_{tot} = (1+ g_{a\gamma} <a\chi> {\omega \over \Omega_*}) \,  \vec S_{BZ} \ ,
 $$
 where the Luminosity of the Blandford -Znajek process \cite{BZ} is
 $$
 L_{BZ} \equiv -{\Omega_* \over \beta_p} (\omega - \Omega_*) \bar r^2{B_p \vec B_p \over 4\pi c} \, 2\pi(1- \cos \theta_o) r_g^2\ ,
 $$
 ($\theta_o$ being the opening angle of the ejection).
 And the power of axion emission stems from the extraction of some small amount of the BZ-power :
 $$
 L_a = \eta_a L_{BZ}   \quad \eta_a \sim \, g_{eff}  {h^2 \over \ell_a^2} \, \chi_0^2   \ .
 $$

When the energy flux of Blandford-Znajek crosses a gap containing axions, it concede a small fraction of power to the axion population, $\langle a \chi \rangle < 0$ (see \cite{GP2}).
This increases the power emission of axions by a factor 2 approximately. Could it be measurable? This process makes sense if $h \simeq \lambda < r_g$ significantly but not too
much smaller, however, as previously mentioned, even if quite narrow these gaps are quite numerous and their number of order 10 compensate their small width, and $\chi_0 < 1$ but not too much smaller.
This could be detected in the surroundings of AGNs and quasars.\\

\subsection{Efficiency of the process}

There are two mechanisms to generate abundant ejection of ULAs in the vicinity of the horizon of a Kerr Black Hole doted of a magnetosphere in the gaps distributed between $r_g$ and $r_0 \simeq 2r_g$ : a direct axion wave emission and an entrainment in the relativistic jet produced by Blandford-Znajek mechanism. The ULA ejection is relativistic but not confined, so that it forms a relativistic wind in radial expansion. The short size of the gaps, $h \sim 0.01-0.1 r_g$, would give rise to short scale ripples; however because $h$ is much smaller than the Compton length $\ell_c$, they are smoothed out. So finally one expects that the conversion parameter $\eta_a \sim 10^{-6}-10^{-5}$.

An important remark is that we considered conservative estimate of the intensity of the magnetic field on the Black Hole horizon, $B_*$, based on the equipartition magnetic field carried by the accretion disk. Actually recent numerical simulations revealed that $B_*$ can be much larger \cite{TCH} \cite{NAR}; therefore both the Blandford-Znajek energy flux  and the axion energy flux are larger. It is useful to have in mind some scaling laws when discussing the efficiency of axion emission in terms of $M_*, B_*, m_a$, the energy $\Lambda$ kept constant. These are the following :
\begin{eqnarray}
g_{eff} & \simeq & 22  \left({1 eV \over \Lambda}\right)^4 \left({B_* \over 10^4 G}\right)^2 \\
\eta_a & \simeq & 2 \times 10^{-3} \left( {1 eV \over \Lambda} \right)^4 \bar P_{esc} {r_g^2 \over \ell_c^2} \left({B_* \over 10^4 G}\right)^2 \\
L_a & = & \eta_a L_* \propto {m_a^2 \over \Lambda^4} M_*^4 B_*^4 \ .
\end{eqnarray}

The empiric law $B_* \sim 10^4 G (M_* / 10^8 M_{\odot})^{-1}$, derived from the equipartition assumption between the pressure  at the inner radius of the accretion disk and the magnetic pressure, that works fairly good for $M_* \leq 10^8 M_{\odot}$, leads a constant value of $L_a$, nevertheless  is an underestimate for higher KBH mass. This is due to the arrested accretion regime around supermassive KBH, where the magnetic field grows beyond its equipartition value \cite{TCH} \cite{NAR}. The equipartition law would make the BZ-luminosity to saturate at $10^{44}-10^{45} erg/s$. This seems to be fairly correct and the axion luminosity dramatically decline for KBH mass larger than $10^8 M_{\odot}$, but not so much;  later on we will examine the estimates for the KBH of Messier 87 which has a mass of $2.4 \times 10^9 M_{\odot}$.
Let us continue to investigate the consequences of the equipartition assumption. Then
$$
L_a \sim \frac{(10^{13} cm)^4}{\ell_c^2 r_g^2} \times 10^{39} erg/s \ .
$$
for cosmological constraints and Lyman-alpha observations, the axion mass is constrained to be in the intervalle $10^{-20} - 10^{-18} eV$ and thus $10^{13} cm \leq \ell_c \leq 10^{15} cm$. The axions of mass $10^{-18} eV \, or \, \ell_c = 10^{13} cm$ are the most efficiently generated axions. The axion luminosity is then
$$
L_a \sim \left(\frac{10^{13} cm}{r_g}\right)^2 \times 10^{39} erg/s .
$$
For a KBH mass smaller than $10^8 M_{\odot}$, the axion luminosity is larger than $10^{39} erg/s$; this would be the case of intermediate black holes. The axion luminosity would decline for masses larger than $10^8$ solar masses, which leads to $r_g > 10^{13} cm$; this will be rediscussed with the case of M87 later on.\\

As for the lowest axion mass to be considered, namely $10^{-20} eV$ and thus $\ell_c = 10^{15} cm$, they can be copiously emitted with a luminosity of  $10^{39} erg/s$ by KBH of mass $10^6 M_{\odot}$. If the Black Hole in the nucleus of Milky Way would rotate sufficiently rapidly and  be threaded by an intense magnetic field, it would generate axions.

Let us come back to the case of M87; which is the closest AGN located at $18 \, Mpc$ from Earth. Its Black Hole environment have been recently mapped \cite{EHT}, which revealed the central dark zone and the magnetic field lines; the estimate of the Black Hole mass is quite precise with $M_* \simeq 2.4 \times 10^9 M_{\odot}$ and thus a gravitational radius $r_g \simeq 3.6 \times 10^{14} cm$. Its magnetic field is estimated as $1-30 \, Gauss$ (equipartition estimate) and the luminosity of the radio jet is between $10^{42}-10^{43} erg/s$. and the reference luminosity $L_*$ is slightly larger, which leads to another estimate of magnetic field : $B_* \simeq 50 \, Gauss$, which is larger than the equipartition estimate. The effective coupling parameter $g_{eff} \sim 3 \times 10^{-2}$, weaker than for $B_* = 10^4 Gauss$. Now the conversion factor :
$$
\eta_a = \, g_{eff} \, \chi_0^2 {h^2 \over \ell_c^2} \sim 3 \times 10^{-6} {r_g^2 \over \ell_c^2} \sim 4 \times 10^{-3} \ .
$$
That conversion factor is much better than that of a typical Black Hole of $10^8 M_{\odot}$, because of the increase of the gravitational radius and despite of the weaker magnetic field but less diminished than with the equipartition assumption. Finally the luminosity of the axion wind of M87 is quite high with
$$
L_a = \eta_a L_* \sim 4 \times 10^{-3} L_{*}  \sim 4 \times(10^{39}-10^{40}) erg/s \ ,
$$
a little larger than the result obtained with fiducial parameters.

\section{An axion wind in a state of a Bose-Einstein condensate.}

As a consequence both classical and quantum predictions of the emission power of axions  are the same \cite{GP1}.
It turns out that this generation of axions creates large Bose-Einstein Condensates. Indeed the ULA density reaches a very high value, which satisfies the criterium for a very dense  BEC. : Indeed consider a gap located above the Black Hole horizon at $r \gtrsim r_g$ and of width $h = 0.1- 0.01 \, r_g$. A gap is highly populated with $n_a (\lambda_a/ \bar \gamma)^3 > 1$ ; then the number
of axions generated in such a gap, according to eq;(...) , is $N_{a \mid gap} \sim (1-10^{-3}) \times 10^{69}$ and the density of axions is $n_{a \mid gap} \sim (1-10^{-2}) \times 10^{31} cm^{-3}$. The number of axions inside a sphere of radius equal to the Compton length measures the level of condensation ${\cal N}_a \equiv n_a \ell_c^3$, and for a set of non relativistic particles, the de Broglie length is larger than the Compton length; so ${\cal N}_a \gg 1$ is a good indicator that the set of bosons is in a Bose-Einstein condensate state. This is the case with the Compton length of the considered axions $\ell_c \sim 10^{13} cm$. Thus
$$
{\cal N}_{a \mid gap} \sim 10^{70} , \quad {\cal N}_{a} \sim 10^{41} \quad at \, 1 Mpc, \quad {\cal N}_{a} \sim 10^{14} \quad at \, 1 Gpc \ .
$$
${\cal N}_a$ is so large and the quantum levels so tight that the condensate does not only concern the ground state but also a significant bunch of excited states and so over a large distance covering the AGN halo.
Therefore the axion wind is governed by a single gigantic Schr\"odinger wave \cite{WK}, which has a peculiar hydrodynamics and the wind gushed out near the Black Hole horizon spreads in the host galaxy and its halo without being fragmented.\\

{\it The escape probability}\\

All the axions generated in the compact region of the KBH are not escaping because of the conversion process of axion into gamma rays. In that source we can estimate the conversion probability with the recent result  derived by Wang \& Ma \cite{WM} involving the effect of the plasma through the parameter q which is a kind of wave number :
$$
q \equiv {\vert m_a^2 - \omega_p^2 \vert  \over \epsilon_{\gamma}} \quad \rm{or} \quad {1\over \hbar c} {\vert m_a^2 - \hbar^2 \omega_p^2 \vert  \over \epsilon_{\gamma}} \quad (usual \, units) \ ,
$$
where $\omega_p$ is the plasma frequency, $\epsilon_{\gamma}$ the energy of the emitted photon; $L$ is the length of interaction with the magnetic field (here $L \simeq 2 r_g$).
$$
P_{a \rightarrow \gamma} \simeq (1/4) (g_{a \gamma} e B L)^2 \left(\frac{\sin qL/2}{qL/2} \right)^2 \ ,
$$
The left factor is a sizable fraction of unity, for $g_{a\gamma} \sim 10^{-20} (eV)^{-1}$ and $eB_*r_g \sim 10^{20} eV$.
Hence that probability is close to unity when $q = 0$, i.e. if there is a resonance $\omega_a = \omega_p$. But the $e^+e^-$-plasma in the source of axions is quite dense \cite{GH} and has a frequency $\sim 10^{10} Hz \simeq 0.6 \times 10^{-4} eV$ very much higher than the axion frequency ($\leq 10^{-18} eV$). Thus the wave number $q \simeq (\omega_p) / c) (\hbar \omega_p / \epsilon_{\gamma}) \simeq 10^{-14} eV$ for a photon emission at $1MeV$. the wave number q is small, however the product $qr_g \simeq 5 \times 10^3$ for a MeV emission. Therefore the product $qr_g$ being very large, the conversion process is negligible and the escape probability of the axions from their source is close to unity. Now there is a non negligible emission of TeV photons with $qr_g \sim 1$, however they are absorbed by pair production in the compact region ($r<100 r_g$) \cite{GH}.\\

In the previous parts of that work (\cite{GP1} \cite{GP2}), we found the luminosity of the ULA emission $L_a = \eta_a L_*$ with $\eta_a = 10^{-6} - 10^{-5}$, so that the ULA emission could reach $\sim10^{39} erg/sec$, which implies the production of a wind.  Despite having a source close to the horizon of the central Black Hole, this flow does escape from its strong gravitation effect, since at its origin at $r_0 \simeq 2r_g$ the gravitation effective potential $\alpha c^2 \simeq c^2/2$, whereas, with a Lorentz factor $\gamma_w \sim 10$, the kinetic energy of the wind axions is $(\gamma_w -1) m_a c^2 \gg \alpha m_a c^2$ ($\gamma_w \gg 1 > \alpha$); we note also that the axions in their wave bottom are strongly trapped. So the source of that emission is above the horizon of the KBH between $r_g$ and $r_0 \simeq 2r_g$ and an escaping axion wind is generated with mass flux
$$
\dot M_a = L_a/c^2 \sim (10^{-16} -10^{-15}) M_{\odot} sec^{-1}  .
$$
The initial energy density of the axion wind is :
$$
\rho_{a \mid 0} = {\eta_a L_* \over 4\pi r_g^2 \gamma_a c} \simeq {\eta_a \over \gamma_a} B_*^2 \sim (10^2-10^3) erg/cm^3 \sim 2 \times (10^{21}-10^{22}) M_{\odot}/ (Mpc)^3 \ ,
$$
for the reference values of $B_*$ and $m_a$. For other values of both parameters, the scaling rule is $\rho_{a\mid 0} \propto m_a^2B_*^4$. In the case of M87, $\rho_{a\mid 0} \sim 1 erg/cm^3$.

Then, the wind being unconfined, its density decreases like $r^{-2}$ :
$$
\rho_a = \rho_{a \mid 0} {r_0^{2} \over r^2} \ .
$$
The cosmic critical density being $\rho_{crit} \simeq 3 \times 10^{11} M_{\odot}/(Mpc)^3$ \cite{PM},  the initial $\rho_{a\mid 0}$ is $10^{10}$ times larger than the critical cosmic density, Then the axion density $\rho_{a}$ declines down to $\rho_{crit}$ at $r_{crit} \sim 10^{5} r_g \sim 1 pc$. It reaches $\rho_{CDM} = \Omega_ch^2 \rho_{crit}$, where  $\Omega_ch^2 \simeq 0.114$ \cite{PM}, at $r_{CDM} \sim 3 r_{crit}$. The baryonic density is $\rho_b \simeq 4 \% \rho_{crit}$, so that the axion density becomes smaller than the cosmic baryonic density for $r > 1 kpc$. But the baryonic gas density inside a normal galaxy is much higher $\sim 10^{17} M_{\odot}/(Mpc)^3$; so that the axion wind is immediately confronted with the baryonic gas in its host galaxy and again with the cosmic baryonic gas in the halo; CDM mass density is more important but no classical interaction occurs with it; another quantum interaction will be considered in section 6. Now an intriguing circumstance is that the axions emitted by the nucleus are still in a condensate state at Mpc scale, $n_a\ell_c^3 \sim 10^{46}-10^{48}$ and even at Gpc scale where this quantity is $\sim 10^{14}$ !..
We note that the axion wind undergoes self-gravitation inside a volume of a few parsec radius. Beyond, for, say, $ r > 100$ parsec, the other components, CDM contribution and baryonic contribution dominate gravity, the CDM contribution being dominant in the halo.\\

During the age of Universe ($\sim 3 \times 10^{17} sec$), the axion wind, if it would not be dissipated, would reach a distance of $\sim 3 \, Gpc$. So we can estimate that the density of ULAs, which decline in $1/r^2$  in a halo of $(1 \, Mpc)^3$ can reach a sizable fraction of the density of Cold Dark Matter :
$\rho_{a} \sim 4 \times 10^{10} M_{\odot}/(Mpc)^3$, whereas the critical density is $\rho_c \simeq  3 \times 10^{11}  M_{\odot}/(Mpc)^3$ \cite{PM}.

However, as observed today, the active galaxies are not numerous enough in the Universe to significantly modify the cosmological parameters. That secondary generation of axion Dark Matter could be more significant at earlier stage of the expansion of Universe with a larger fraction of the new born galaxies would be in an active regime. This might contribute in explaining the cosmological results of DESI experiment \cite{DESI}.\\

{\it Expansion velocity.}\\

The axion wind can be considered as a coherent wave having a wave number
$$
k_0 = \pm {\omega_0 \over c} \sqrt{1- {\omega_a^2 \over \omega_0^2}}
$$
and is unavoidably superluminal :
$$
\upsilon_{\phi} = {\omega_0 \over k_0} = c \left(1- {\omega_a^2 \over \omega_0^2} \right)^{-1/2} \ ,
$$
whereas the group velocity is sub-relativistic as expected, since $v_{\phi} v_g = c^2$ :
$$
\upsilon_g = c  \left(1- {\omega_a^2 \over \omega_0^2} \right)^{1/2} \ .
$$
That group velocity is the wind propagation velocity.

Actually the process is not so mono-energetic, because the gaps are spatially distributed, some disappear whereas others appear, the oscillation frequencies are not exactly the same. So it is relevant to consider some distribution of resonant frequencies and thus a dispersion of the phase and group velocities. Thus a relativistic axion flow is expected with a Lorentz factor $\gamma_0 = \omega_0/\omega_a$ having a dispersion
$\Delta \gamma_w = \Delta \omega_0/ \omega_a$. Therefore through out the paper we will assume that the axions in the wind have a Lorentz factor $\gamma_a \simeq \gamma_0 \equiv \gamma_w \sim 10$, with a small non relativistic dispersion.

We note that this axion wind is in rotation; the effect of rotation in a superfluid is a very tricky problem that we will not address.

\section{Quantum hydrodynamics and thermalization.}

 Up to large distance from the nucleus, the axion wind dynamics can be described by a single quantum wave, and because it has a relativistic motion, that quantum wave can be factorized as a Klein-Gordon factor times a Schr\" odinger wave that describes the non-relativistic deviations to the global relativistic expansion at $\upsilon_a = \upsilon_g \simeq c$
 with a global Lorentz factor $\gamma_0 \simeq \omega_0/\omega_a$ having some slight dispersion
 $\Delta \omega_0/ \omega_a < 1$ :
 $$
 a(x,t) = \exp[i(\vec p \cdot \vec x - \epsilon t)/\hbar] \psi(x,t) \ ,
 $$
 with $\vec p = m_a \gamma_a \vec \upsilon_a$ ($\gamma_a \simeq \gamma_0$),
 the Schr\"odinger factor being of the form $\psi(x,t) = \sqrt{\rho(x,t)} \exp[i\theta(x,t)]$,
 making the transition from the Klein-Gordon equation to the Schr\"odinger one introduces a mass $\gamma_0 m_a$ and thus the Compton length is divided by $\gamma_0$.\\

The hydrodynamical description in the co-moving frame of the relativistic wind stems from the  use of a system of two coupled equations for $\rho$ and $\theta$, by writing a Schr\"odinger wave as $\psi(x,t) = \sqrt{\rho} \exp(i\theta(x,t))$ \cite{HOTW} \cite{WK}. Then one changes the dimensions of both fields such that $\rho$ becomes a mass density : $\gamma_0 m _a \vert \psi \vert^2 \mapsto \rho$ and $\theta$ gets the role of an hydrodynamical potential $\phi$ such that $\theta \mapsto \phi = \omega_a \ell_c^2 \theta$.
The velocity in the co-moving frame, supposed to be non-relativistic, is $\vec \upsilon = \nabla \phi$. Then the Hamiltonian is changed as a irrotational hydrodynamical Hamiltonian with a peculiar enthalpy density :
$$
H = \int dx^3 [{1\over 2} \rho (\nabla \phi)^2 + {c^2 \over 8} {\ell_c^2 (\nabla \rho)^2 \over \rho} + \rho V] \ .
$$
 the system is the following:
\begin{eqnarray}
\partial_t \rho = {\delta H \over \delta \phi} & =  & - {\rm div} \, (\rho \nabla \phi) \\
\partial_t \phi = - {\delta H \over \delta \rho} & =  & - [{1\over 2} (\nabla \phi)^2 + h + V] \ .
\end{eqnarray}
The scalar field $V$ is the gravitational potential and $h$ is a sort of enthalpy density such that
$$
h \equiv {c^2 \over 8}{\ell_c^2 \over \gamma_0^2} {\delta \over \delta \rho}  \int {(\nabla \rho)^2 \over \rho} d^3x = {1\over 8}{c^2 \ell_c^2 \over \gamma_0^2} \left({(\nabla \rho)^2 \over \rho^2} - 2 {\Delta  \rho \over \rho}\right) = {1\over 2} c^2 {\ell_c^2 \over \gamma_0^2} \frac{\Delta \sqrt{\rho} }{\sqrt{\rho}} \ .
$$
The pseudo quantum pressure is such that $\delta P = \rho \delta h$; that pressure is responsible for the quantum spreading. But, according to Messiah \cite{AM}, the momentum fluctuation does not grow nor decay and the flaring out does not imply a slowing down, whereas the fluctuation of the particle position increases like $t^2$ like a super-diffusion.\\

In absence of perturbation the axion wind has a density decreasing like $1/r^2$. Then its enthalpy density vanishes. However there is a natural production of ripples in the wind due to the dispersion of the initial 4-velocity of the wind. Indeed, close to the KBH horizon, the axion generation is not produced by a pure oscillation  in a gap; moreover there are several gaps which do not oscillate with exactly the same frequency.  The dispersion of these frequencies implies a dispersion of the resonant velocities. So there is a mean 4-velocity of the wind with some non relativistic dispersion. In the co-moving frame with the Lorentz factor $\gamma_0$, there is a dispersion of the initial velocity $\Delta \upsilon$ around the value zero. Only perturbations produce enthalpy variations. This is what happens when one considers the effect of initial dispersion of the velocity of the wind. However we will not develop these issues in this paper.\\\

{\it Relaxation through quantum interferences ?}\\

In cosmological studies under the assumption that Dark Matter is mainly made of ULAs, the quantum interferences between ULA flows in a BEC state can play a dynamical role through their contributions to Newton equation.  Their contributions are $\propto G$ and have been successfully considered for the thermalization of Dark Matter \cite{ERK}.
Indeed let us consider two axion streams of momenta $\vec p_1$ and $\vec p_2$ with respect to an arbitrary frame, which undergo non-relativistic quantum fluctuations and are in Bose-Einstein condensate. Then each stream is described by a wave function of the form $a_{\alpha}(\vec x,t) = \exp[i(\vec p_{\alpha} \cdot \vec x - \varepsilon_{\alpha} t)/\hbar] \tilde \psi_{\alpha}(\vec x, t)$, with $\varepsilon = \sqrt{m^2+p^2}$ in reduced units, and where $\tilde \psi_{\alpha}$ is the Schr\"odinger wave function of the axion $\alpha$ of mass $\gamma_{\alpha} m$, $m$ being the same mass for both axion streams and $\omega_{\alpha} = \gamma_{\alpha} m/\hbar$ (actually there is no mistake in incorporating the relativistic momentum of the wind in the phase).\\

Let us consider two flows of axions having the same mass that differ from their momenta $\vec p_1$ and $\vec p_2$.
$$
a_{1,2} = \sqrt{\rho_{1,2}} \exp[{i\over \hbar}(\vec p_{1,2} \cdot \vec x - \epsilon_{1,2} t) +i \tilde \theta_{1,2}] \ ,
$$
where $\tilde \theta_{1,2}$ represents the Schr\"odinger phases of both the wind and Dark Matter in their co-moving frame.
The densities $\rho_{1,2}$ are normalized such that they represent the mass density of each stream and we consider the generation of gravitational field due to streaming axions according to the Poisson equation
${\rm div} \, \vec g = -4\pi G \vert a_1 + a_2 \vert^2$. That density contains a quantum interference contribution  :   $\vert a_1 + a_2 \vert^2 = \rho_1+ \rho_2 + 2 \sqrt{\rho_1 \rho_2} \cos (\theta_1- \theta_2)$; and the phase difference responsable for the interferences introduces a gravitational interaction due to the difference of momenta (see \cite{ERK} for a more extensive study) :
$$
\theta_1-\theta_2 = {1\over \hbar} [(\vec p_{1} -\vec p_2) \cdot \vec x - (\epsilon_1-\epsilon_2) t] + \tilde \theta_1-\tilde \theta_2 \simeq {1 \over \ell_c}(\Delta \vec u \cdot \vec x -\Delta u^0 ct )  + ... \ .
$$
where $\vec u \equiv \gamma \vec \upsilon /c$, $\Delta \vec u = \vec u_1 - \vec u_2$ and $u^0 = \gamma$, $\Delta u^0 = u^0_1-u^0_2$; here the Compton length $\ell_c$ is the proper one, and we assume that the non relativistic part of the phases are negligible compared to the relativistic contributions.
The interferences introduced a gravitational interaction in Poisson equation :
\begin{equation}
\Delta V = 4\pi G [\rho_1 + \rho_2 +  \rho_b + 2 \sqrt{\rho_1 \rho_2} \cos \left({1 \over \ell_c}(\Delta \vec u \cdot \vec x -\Delta u^0 ct ) \right)] \ .
\end{equation}
When introduced in the equation of motion, at distance larger than the Compton length, a kind of drag force in $\vec p_1- \vec p_2$ is generated and is expected to produce the thermalization.

Now let us specify our calculation. Our flow of interest is the relativistic axion wind, supposed isotropic and of density declining in $r^{-2}$ and the other one is the axionic Dark Matter, that we suppose homogeneous and at rest with respect to the galaxy frame. Therefore The radial gravitational field due to those quantum interferences is such that :
$$
{1\over r^2} {\partial \over \partial r} r^2 g_{int} = -8\pi G \sqrt{\rho_a \rho_{dm}} \cos(\gamma (\beta_a r - ct)/\ell_c) \ ,
$$
where $\beta_a$ is the radial velocity of the wind, $\ell_c$ is the proper Compton length of the ULA axions. The density of the axion wind is such that $\rho_a = \rho_{a \mid 0} r_0^2/r^2$, $r_0$ being the size of the source of the wind $r_0 \gtrsim r_g$. We define a frequency $\nu_{int}$ characteristic of the interaction by the interferences such that :
$$
\nu^2_{int} \equiv 4\pi G \sqrt{\rho_{a \mid 0} \rho_{dm}} \ ,
$$
then the gravitational field component generated by the interferences during a large part of the wind dynamics is approximately governed by the following approximation for $r \gg \ell_c$ :
$$
g_{int} \simeq -2\nu_{int}^2 {r_0 \over r}  \, \frac{\sin((\beta_a r -  ct) \gamma_a/\ell_c)}{u_a/\ell_c} \ .
$$
This makes sense only if the involved Dark Matter is also in a BEC state, otherwise, notably decoherence rapidly destroys the interferences.\\

{\it At which distance is that quantum interaction effective ?}\\

The slowing down is governed by the relaxation equation :
\begin{equation}
u_a \dot u_a = -2 \nu_{int}^2  {r_0 \ell_c \over rc} \, \sin[(\beta_ar -ct) {\gamma_a \over \ell_c}] \ .
\end{equation}
First we look at the relativistic regime of relaxation : $u_a \simeq \gamma_a$, $\beta_a \simeq 1$, $r(t) \simeq r_0 + ct$. Then the relativistic slowing down is governed by :
\begin{equation}
\gamma_a \dot \gamma_a = -2 {c \over r} {r_0 \ell_c \over \lambda^2_{Ja}} \sin[(\beta_ar -ct) {\gamma_a \over \ell_c}] \simeq -2 {c \over r} {r_0 \ell_c \over \lambda^2_{Ja}} \sin[{r_0 \over \ell_c} \gamma_a] \ .
\end{equation}
The length $\lambda_{Ja}$ is a kind of Jeans length such that
$$
\lambda_{Ja}^2 \equiv c^2/ \nu_{int}^2 = {c^2 \over 8\pi G \sqrt{\rho_{a \mid 0} \rho_{dm}}} \ ,
$$
Assuming $\ell_c = r_0$, one obtains the simple differential equation:
$$
{\gamma \over \sin \gamma} d \gamma = -2 {r_0^2 \over \lambda_{Ja}^2} {dr \over r}
$$
That Jeans length $\lambda_{Ja} \sim 0.1 Gpc$, whereas $r_0$ and $\ell_c$ are of order $r_g \simeq 10^{-5} pc$. The Lorentz factor of the wind does not decline faster than allowed by the following inequality:
$$
\gamma(r_0) -\gamma(r) \leq 4 {r_g^2 \over \lambda_{Ja}^2} {\rm Ln} \, {r \over r_0} \ .
$$
Thus the isotropisation of the axion population is impossible in the galactic halo; it occurs at very large scales only, beyond several $100 \, Mpc$, where the KBH axions are mixed with the Dark Matter axions. This differs mostly with cosmology because the source of the wind is at a size $r_0$ extremely small ($\sim r_g$) compared to the Jeans length $\lambda_{Ja}$. Consequently, this process cannot generate acoustic waves either.\\

\section{Generation of baryonic acoustic waves by the axion wind}

First of all, we have to bear in mind that the life-time of ULAs is very long; indeed their decay rate in two photons is very small. The thermalization of an axion cloud through self-interaction by the nonlinearity in $\phi^4$ is a very long process also. If we would not consider the axion flows organized as a BEC, all the classical gravitational two body interactions would develop relaxation $\propto G^2$ and would fail to operate over a time shorter than the age of Universe. This concerns scattering off the population of stars, off the baryonic gas, and off the axionic Dark Matter supposed to be populated by the same ULA than the wind axions. We will examine
the collective scattering later on.

\subsection{Preliminary settings}

In the same spirit as the recent investigations developed in cosmology, called BAO (Baryonic Acoustic Oscillations) \cite{BAO}, one could expect to find similar features in the host galaxy of AGNs and in its halo and beyond, through the gravitational interaction of the axion winds with the ambient baryonic matter, which would require comparisons with numerical simulations. They could be diagnosed by the correlations of Lyman alpha forests and the method developed in \cite{XRI}. This is the motivation of the investigations presented in this section, where the presentation of a fast instability could hopefully lead to an observable generation of acoustic waves.\\

{\it The gravitational Tcherenkov-Vaviloff effect.}\\

Before doing more complex calculation in spherical geometry, let us look at the elementary gravitational version of the Tcherenkov-Vavilov process in cartesian geometry, using Fourier transform.
The acoustic waves have a frequency such that $\omega_s = kc_s$ and the axion motion generates a vibration of frequency $\vec k \cdot \vec \upsilon_a$, because of the strong Doppler effect. The resonance between these two vibrations occurs when  $kc_s= k \upsilon_0 \cos \alpha$, so that $\cos \alpha = c_s/\upsilon_0$.
Since $\upsilon_0 \gg c_s$ the angle $\alpha$ is almost perpendicular to the axion velocity, and $k_{\ell} / k = c_s/\upsilon_0 \ll 1$ whereas the transverse component of the wave vector $k_t/k \simeq 1$. We will adapt similar constraint of the resonance in the realistic description of the excitation of acoustic waves by a spherical axion wind further on.\\

{\it The acoustic wave generation by the travelling axion fronts.}\\

The analysis now is done in spherical coordinates and instead of using Fourier transform, we expand in spherical harmonics. Whereas the axion density decreases in $1/r^2$, the baryonic background is supposed homogeneous.
The generation of acoustic waves by the axion wind is governed by the following equation, where $\Delta_{S_2}$ represents the angular laplacian operator in spherical coordinates
($\theta, \varphi$) and the function $\mu$, such that $\delta \rho_a = \rho_{a\mid 0} (r_0^{2} /r^2) \mu$,  represents the various density ripples occurring in the axion wind, the function
$g$ is the reduced version of $\delta \rho_b$ ($\delta \rho_b = \rho_{a \mid 0} \nu_b^2 r_0^{2} g$) :
\begin{equation}
[r^2 \partial_t^2 - c_s^2 (\partial_r r^2 \partial_r + \Delta_{S^2})] \, g = \mu(r-\upsilon_0 t, \theta,\varphi) \ .
\end{equation}
We expand the functions $\mu$ and $g$ in spherical harmonics (the eigenvalues of $\Delta_{S^2}$ are $\ell(\ell+1)$ independently of the azimuthal number $m$, $-\ell \leq m \leq +\ell$ and make a temporal Fourier transform at each radial position.
Then having introduced the function $g(r, \theta, \varphi; t)$ and expanded it in spherical harmonics also, then we make a
temporal Fourier transform of both expansion components, finally
we obtain the relation between the expansion components. \\

According to the previous paragraph, we seek for a Tcherenkov resonance such that $\omega_s = k c_s = k_r \upsilon_0$ with wave numbers adapted with the spherical geometry. The wave number $k$ is the largest one with $k^2 \gtrsim k_t^2 = \ell(\ell+1)/r^2$, $k^2 = k_t^2 +k_r^2$ ; $k_r^2$ is much smaller because of the Tcherenkov resonance that implies $k_r^2 = (c_s^2/\upsilon_0^2) k^2$; $k_r^2$ is a measure of the strength of the operator
$\partial_r r^2 \partial_r$ much smaller than the laplacian operator on the sphere $\Delta_{S^2}$, measured by $k_t^2 =  k^2-k_r^2 \simeq k^2$. Therefore $k_t^2 \simeq \ell(\ell+1)/r^2$  and the resonance condition implies $k_r^2 \simeq (c_s^2/\upsilon_0^2) \ell(\ell+1/r^2)$. Actually to be more accurate, we should talk about
a magneto-sonic wave instead of a mere sonic wave, especially in the central region of the host galaxy where the magnetic field is fairly intense.

The function $\mu$ describes both the generation of acoustic waves by the ripples carried by the wind and the back-reaction produced by the acoustic response of the oscillating gravitation field.

It turns out that the generation of acoustic waves by the wind ripples is weak and thus we present the instability produced by the back-reaction.

\subsection{A streaming instability in $G^{1/2}$}

In that subsection we investigate the dynamics of the baryonic acoustic waves through  their resonant gravitational interaction with the axion wind carrying ripples, taking account of the back reaction of the acoustic waves
on the wind, the ripples being somehow self consistently generated by the acoustic waves themselves.
We assume some simplifying assumptions as follows : the baryonic fluid is homogeneous in the host galaxy and in its halo, with different densities and sound speeds, similarly the Dark Matter fluid is homogeneous and thermalized  everywhere. The axion wind has a density declining in $r^{-2}$, a relativistic velocity $\upsilon_0 \simeq c$ and a Lorentz factor $\gamma_0$ of a few tens, and a massive enthalpy $h \ll c^2$, that will be neglected in the study of the destabilization of baryonic acoustic waves. The new effect stems from the back reaction of the generation of acoustic waves on the axion wind, which suffers
a supplementary gravitational perturbation due to the acoustic wave, that plays the same role than ripples but at a specified frequency, and we expect that the back reaction will drive an instability.\\

{\it The growth rate}\\

In this paragraph we investigate the development of an instability involving a Tcherenkov resonance in a collisionless regime. We neglect the Dark Matter perturbation and linearize the motions of both wind density $\delta \rho_a$ and baryonic density $\delta \rho_b$ :
\begin{eqnarray}
c^{-1} \gamma_0 \partial_t \delta \rho_a + {\rm div} \, (\bar \rho_a \delta \vec u + \vec u_0 \delta \rho_a) & = & 0 \\
\bar \rho_a \gamma_0 (c^{-1} \gamma_0 \partial_t \delta \vec u + \vec u_0 \nabla \delta \vec u)& = & \gamma_0 \bar \rho_a \delta \vec g \ ,
\end{eqnarray}
where we assumed a large enough distance $r$ such that $r^2 \gg \ell (\ell +1) \ell_c^2$ allowing to neglect the enthalpy variation compared to that of the convective term and thus
\begin{eqnarray}
\gamma_0 (\partial_t + \vec v_0 \cdot \nabla)^2 \delta \rho_a & = & c^2 {\rm div} (\bar \rho_a \delta \vec g) + ... \\
(\partial_t^2 - c_{s}^2 \Delta) \delta \rho_{b} & = & \bar \rho_{b} {\rm div} \, \delta \vec g \ ,
\end{eqnarray}
where  $c_s$ is the sound speed of the baryonic gas (or its magneto-sonic velocity). The coupling between these fluids stems from Newton equation :
$$
{\rm div} \, \delta \vec g = -4\pi G (\delta \rho_a + \delta \rho_b) \ .
$$

In the host galaxy, the baryonic fluid is much more massive than the Dark Matter fluid. Therefore we consider the coupling between the wind fluid and the baryonic one, where $k^2 r^2 \simeq l(l+1)$ and $\vec k \cdot \vec \upsilon_0 = k_r \upsilon_0$ with $k_r$ determined by the resonance condition as before: $k_r^2r^2 \simeq \ell(\ell+1) c_s^2/\upsilon_0^2$  :
$$
[(\omega^2 - k^2 c_s^2) + \nu_a^2][ (\omega - \vec k \cdot \vec \upsilon_0)^2 + \nu_b^2] =  \nu_a^2 \nu_b^2  \ .
$$
We note $\nu^2_b \equiv 4\pi G \bar \rho_b$ and $ \nu_a^2 \equiv 4\pi G \bar \rho_a/\gamma_0$. We look at baryonic oscillations of wave lengths shorter than the respective Jeans lengths ($c_s/\nu_a$ and $c_s/\nu_b$). Thus for the calculation of the instability, we will use of two small parameters : $\epsilon_a \equiv \nu_a^2 / \gamma_0 k^2c_s^2$ and $\epsilon_b \equiv \nu_b^2/ k^2c_s^2$. We note that $\epsilon_a$ is a constant $(\nu_{a\mid 0}^2 r_0^2 / \gamma_0 \lambda_{Ja}) \times \ell(\ell+1)$. We assume $\rho_b$ homogeneous in each domain of investigation with three different values in the host galaxy, its halo and its surrounding. We anticipate that the maximum of instability occurs at Tcherenkov resonance ($\vec k \cdot \vec v_0 = k c_s$) and that the acoustic frequency is slightly modified by the gravitational interaction between axions and baryons : $\omega = kc_s(1+z)$ with $\vert z \vert \ll 1$. The number $z$ is the root of a third order polynomial equation:
$$
z^3+ 2z^2 + (\epsilon_a+\epsilon_b)z + 2\epsilon_a = 0 \ .
$$
We have to compare both small parameters $\epsilon_a$ and $\epsilon_b$. In all circumstances, $\epsilon_b \ll \epsilon_a \ll 1$. For $\epsilon_a > \epsilon_b/12$,  one always obtains a root with a positive imaginary part when $\epsilon_b < \epsilon_a$,  which means that the wind triggers an instability. One finds
$$
z \simeq  - {\epsilon_a + \epsilon_b \over 4} + i \sqrt{\epsilon_a} \ .
$$
The acoustic frequency is slightly shifted :
$$
\omega_s = kc_s (1- {1\over 4} {\nu_a^2/\gamma_0 + \nu_b^2 \over k^2c_s^2}) \ ,
$$
and the growth rate independent of $k$ :
$$
g \simeq \nu_a = {\nu_{a \mid 0} \over \gamma_0}  {r_0 \over r} \ .
$$
Over the age of Universe $t_U = 3 \times 10^{17} s$, with $\nu_{a\mid 0} \simeq 2 \times 10^{-12} sec^{-1}$, $r_0 \simeq 0.3 \times 10^{-5} pc$ and $\gamma_0= 10$, the baryonic oscillations achieved an amplification by a factor $gt_U \simeq 0.6 pc/r$. Therefore the acoustic waves undergo a amplification over a few parsecs only and with a sound speed $c_s \sim 10^{-6} c$ (in the case of a magneto-sonic wave the velocity would be much larger), they propagate over a distance $c_st_U \sim 3 kpc$. Crossing the region of acoustic wave generation in $10^8s$, the axion wind is weakly affected by that acoustic emission and then it continues its travel across the halo and the surrounding intergalactic medium. Around $10 Mpc$, it mixes with other axion winds and they possibly thermalize together at that scale \cite{ERK} and increases the Dark Matter density.\\

About collisional effects that we have not taken into account, in a normal galaxy, where the baryonic density is $n_b \sim 0.1 \, atoms/cm^3$, the mean free path $\lambda_{mfp} \sim 0.1 pc$. In an elliptical active galaxy, the baryonic density is much weaker and $\lambda_{mfp} \gg 1 pc$. Therefore the central region of the host galaxy of KBH where acoustic waves are generated, the baryonic collisions are negligible.

\subsection{Scattering}

At the scales between $100 Mpc$ and $4 Gpc$, there are many active galaxies that are supposed to produce axion winds as well. Therefore the axion winds propagating in that scale range are likely scattered by the baryonic acoustic waves excited by the numerous axion winds. The scattering rate can be roughly estimate as follows. We consider the pitch angle $\alpha$ of an axion momentum deviated from its initial radial direction by  a pitch angle
$\Delta \alpha$  ($\mu \equiv \cos \alpha$) during $\Delta t$; the scattering frequency is then

$$
\nu_{scat} \equiv {\langle (\Delta \vec p)^2 \rangle \over 2p_0^2 \Delta t}  = {\langle \Delta \mu^2 \rangle \over 2 \Delta t} = {1 \over \gamma_0^2 \upsilon_0^2} \int_0^{\Delta t} C (\tau) d\ \tau \ ,
$$
where $C(\tau)$ is the correlation function of the gravitational field fluctuations produced by the baryonic density perturbations.  We define the correlation time $\tau_c$ such that
$$
\int_0^{\infty} C(\tau) d\tau = \langle \delta g_r^2 \rangle \tau_c \ ;
$$
this is a measure of the duration of the travel of an axion over a correlation length $\bar \lambda_s$ of the acoustic perturbations. If an axion keeps the velocity $\upsilon_0$ along this travel (in the spirit of Quasi Linear Theory (QLT);  then $\tau_c = \bar \lambda_s/\upsilon_0$; but if the axion is scattered during its travel, the correlation time is not longer than the scattering time $\nu_{scat}^{-1}$ (in the spirit of Resonance Broadening Theory (RBT);
see \cite{ML} and references within), the RBT estimate leads to a faster scattering than the QLT one.\\

Now the Fourier spectrum of $\delta \vec g$ is quite erratic due to presence of many AGNs in the considered volume; then
$$
\langle \delta \vec g(k)^2 \rangle = (4\pi G)^2 \frac{\langle \delta \rho_b(k)^2 \rangle}{k^2} \ .
$$
Then we reasonably assume that the correlation length of the gravitational fluctuations is comparable with the correlation length of the baryonic fluctuations, so that
$$
\langle \delta \vec g^2 \rangle \simeq  (4\pi G)^2  \int d^3k \langle \delta \rho_b(k)^2 \rangle / k^2 \sim \langle \delta \rho_b^2 \rangle \bar \lambda_s^2 \ .
$$
Thus inserting the correlation time, one obtains an estimate of the scattering frequency :
\begin{equation}
\nu_{scat} = {\langle \Delta \alpha^2 \rangle \over 2 \Delta t} = {(4\pi G)^2 \over \gamma_0^2 \upsilon_0^2} \bar \lambda_s^2 \tau_c = \\
{\cal A} {\nu_b^4 \over \gamma_0^2 \upsilon_0^2} \bar \lambda_s^2  \tau_c \ ,
\end{equation}
where we have inserted  the baryonic gravitation frequency $\nu_b$ and the acoustic amplitude factor ${\cal A} \equiv \langle \delta \rho_b^2 \rangle / \bar \rho_b^2$. Therefore we got two possible estimates with two possible correlation time:
\begin{eqnarray}
\nu_{scat} & = & {\cal A} {\nu_b^4 \over \gamma_0^2 \upsilon_0^3} \bar \lambda_s^3    \quad QLT \quad and \\
\nu_{scat} & = &  {\cal A}^{1/2} {\nu_b^2 \over \gamma_0 \upsilon_0} \bar \lambda_s  \quad RBT \ .
\end{eqnarray}

Let us see the efficiency of the QLT scattering when the host galaxy suffers an intrinsic acoustic turbulence so that ${\cal A} \lesssim 1$ independently of any wind instability. Then $\nu^2_b$ is $10^{7}$ larger than its cosmological value, and for $\bar \lambda_s = 100 pc$, then $\nu_{scat} \sim 10^{-30} s^{-1}$ and for $\bar \lambda_s = 30 kpc$ it would be better with $\nu_{scat} \sim 3 \times 10^{-23} s^{-1}$; in both cases the scattering is too weak and this is sufficient to rule out an efficient QLT scattering in galaxies... Now outside the galaxy assuming the cosmological baryonic density,  with $\bar \lambda_s = 1 Mpc$, one finds $\nu_{scat} \sim 10^{-32} s^{-1}$; this is worse... Thus even if there is an instability is strong enough to produce an acoustic turbulence such that ${\cal A} \lesssim 1$, the QLT scattering remains weak and does lead to a relaxation of the wind. This seems so weak that its looks non realistic... \\

Does the RBT assumption allow to get some scattering of the axion wind in a time shorter than the age of Universe? This mainly depends on the unknown correlation length $\bar \lambda_s$. Inside a normal galaxy where $\nu_b^2 = 10^7 \nu_{b \mid cos}^2$,
$$
\nu_{scat} \sim 0.3 \times 10^{-17} (\bar \lambda_s / 300 pc) sec^{-1} \sim t_u^{-1} \ ;
$$
this seems marginally possible, but the transit time across the galaxy is $3 \times 10^{12}s.$, thus the scattering is not fast enough. However for an elliptical galaxies of much weaker baryonic content, this is ruled out.  The negative result for a normal galaxy in interesting because it shows that it is possible to determine the origin of extragalactic axions undergoing gamma-conversion when crossing Milky Way \cite{MEX}. Whereas outside a galaxy $\nu_{scat} \sim 10^{-23} (\bar \lambda_s / 1 Mpc) sec^{-1}$; the scattering is clearly too weak again. That weakness of the axion scattering makes possible an extragalactic astronomy of axion sources.\\

So at this stage, one cannot exclude that the relativistic stream of axion flows be continuing their travel beyond the age of Universe. Nevertheless at large scales there is a mixture of many winds coming from a lot of AGNs. Indeed the density of AGNs in the local Universe is $(1-2)\times 10^{-3} / (Mpc)^3$ \cite{IA} and thus the average distance between AGNs is $\simeq 10 Mpc$.  At $5 Mpc$ from an AGN, the density of axions is decreased by a factor $4 \times 10^{-24}$;  therefore as a source of axions, each AGN is very distinguished and an axion astronomy would be possible.

Even if the axion winds are not scattered off, they should contribute to an excess of Dark Matter in the Universe. Is that explaining the DESI measurements? The numbers of these events associated with active galaxies, is probably not sufficient to modify significantly the contribution of all the galaxies in the local Universe and would probably not modify significantly the cosmological parameters measured by Planck mission \cite{PM}.

\subsection{A special  estimate with the M87 parameters}

The previous estimates were proposed with the roughly usual numbers considered for most AGNs, that we considered as "fiducial". However the closest AGN, the nucleus of the active galaxy Messier 87, displays quite different parameters, which are interesting for our purpose and for future investigations. Its distance from Earth is $16,4 Mpc$ and the mass of the central Black Hole, which has been recently mapped with its magnetic lines \cite{EHT}, is much larger than the fiducial reference we considered, with the considerable value $M_* \simeq 2.4 \times 10^9 M_{\odot}$; the gravitational radius is then 24 times larger with $r_g \simeq 36 \times 10^{13} cm \simeq 10^{-4} pc$ and the magnetic field weaker than fiducial with $B_* = 1-30 \, Gauss$, however not precisely determined, nevertheless roughly compatible with the law $B_* r_g  = constant$. However the luminosity of the radio jet is $L_{jet} = 10^{42}-10^{43} erg/s$ and $L_* \equiv cB_*^2 r_g^2 \gtrsim L_{jet}$, so that the required magnetic field  is larger with $B_* \sim 50 \, Gauss$. However the accretion disk likely undergoes the ``arrested regimes", where the magnetic field at the inner radius dramatically increases  and arrests the accretion flow. Ejections would likely occur during this stage of arrested accretion, due to the strong increase of magnetic field.
Now the effective coupling constant $g_{eff}$,  proportional to $B_*^2$ is weaker, but the conversion factor $\eta_a$ is significantly increased due to the increase of $r_g^2$ by a factor $1300$. Therefore  $g_{eff} \simeq 0.25 \times 10^{-2}$, as for the conversion factor $\eta_a = \, g_{eff}\chi_0^2 h^2/\ell_c^2 \sim 4 \times 10^{-3}$. Thus the axion luminosity is higher : $L_a = \eta_a L_* \sim 40 \times 10^{39} erg/s$, which is $40$ times the fiducial estimate. As for the growth rate of acoustic instability, $\rho_{a\mid 0}$ is smaller than the fiducial estimate by a factor $10^{-2}-10^{-3}$ and the gravitational radius larger by a factor $36$.
Thus the acoustic waves are still excited on several parsec scales. However as we saw in the previous paragraph, the scattering is not efficient to slow down the axion wind that continues its travel at relativistic speed up to a thermalization region at several $100 \, Mpc$ and even crosses Milky Way without possible detection except the gamma emission due to axion conversion into gamma rays. \cite{MEX}!... As for the acoustic diagnosis with Lyman-alpha technique, it should be done in the bulge of M87 at parsec scales.

\section{Discussion}

- The generation of axion particles above the horizon of a massive Kerr Black Hole dressed by a rotating magnetosphere has been proved to be very effective \cite{GP1} \cite{GP2}. The density of the created axions is so large that
it leads to the formation of a Bose-Einstein condensate and it remains so all along the axion wind expansion up to extragalactic distances. If the concept of axion particle is relevant, especially the ULA concept, very promising candidate for Dark Matter in cosmology,  then that generation of axion wind in the vicinity of massive Kerr Black Holes at the heart of Active Galactic Nuclei is unavoidable. Moreover our analysis has been presented with a conservative estimate of the intensity of the magnetic field, i.e. the intensity achieved in the inner part of the accretion disk. However it has been recently shown by numerical simulations that the accumulation of magnetic field by accretion onto a Black Hole can lead to a more intense magnetic field in supermassive Kerr Black Hole when in the so-called MAD regime \cite{TCH} \cite{NAR}. A more intense magnetic field leads to a more energetic and denser axion wind.\\

- The major question that raised in this study is the fate of this wind, knowing that these axions particles have not interaction characteristics, except their mass and only gravitational effects can be envisaged.
 The relativistic expansion of that wind turns out to be difficult to stop because of the weakness of the only possible interaction with other type of particles, namely gravitation. Gravitational interaction with baryonic matter can generate acoustic waves, like in the ``BAO" in cosmology, but its investigation should be done with a different method. Once baryonic acoustic waves are excited by the wind, one may expect some scattering effect and slow down of the axion wind; but it turns out to be very weak and the axion wind propagate without significative loss up to extragalactic distances.\\

 - We have investigated various processes of generation of baryonic acoustic waves through interaction in $G$ and $G^2$ and selected the most efficient one : an instability  mediated by a gravitational Tcherenkov resonance, leading to an instability growth rate in $G^{1/2}$.  That instability is sufficiently efficient and has a growth time shorter than the age of Universe inside a volume of one parsec; therefore baryonic acoustic waves are mostly generated in the bulge of the host galaxy. These acoustic perturbations of the baryonic gas might be revealed by an appropriate Lyman-alpha program. The excited acoustic waves propagate mostly in the transverse direction, exciting all the spherical harmonics. If these acoustic patterns could be observable by a program similar to a BAO program \cite{BAO} but with a different procedure involving an appropriate numerical simulation, or with XRISM program \cite{XRI}, they may hopefully reveal the existence of these intriguing particles.\\

 -We found that the quantum thermalization of the wind with Dark Matter, an order $G$ interaction, cannot occurs before several $100 \, Mpc$. Thus a kind of axion extragalactic astronomy could be envisaged within distances shorter than several $100 \, Mpc$, a volume that contains many AGNs.\\

 - Looking specifically at Messier 87, we found that this powerful AGN, quite close to Milky Way (at a distance $16.4 \, Mpc$), should generate a large amount of ULA axions that would be crossing Milky Way and producing gamma rays by the conversion process \cite{MEX} and also be producing Lyman-alpha acoustic patterns in the central region of the galaxy at parsec scales. Similar diagnosis could be realized with other close AGNs like Centaurus A, 3C273, Markarian 421. Actually there are many AGNs in a volume of $200 \, Mpc$ around Milky Way, with an average distance between them of $10 \, Mpc$, also possibly generating axion winds and acoustic waves. Therefore two opportunities of axion wind diagnostics  are suggested in this work : making evidence of acoustic patterns in the center of active galaxies and developing extragalactic axion astronomy within say $200 \, Mpc$ with appropriate detectors based on axion-$\gamma$-photon conversion.\\

\acknowledgments

The author feels very grateful for fruitful discussions with his former collaborators: Gilles Henri, Jonathan Ferreira, Martin Lemoine and Beno\^it Cerutti.


\end{document}